\documentstyle[prl,aps,floats,graphics,twocolumn]{revtex}

\begin{document}

\noindent
{\bf Houdayer and Martin reply:}
\bigskip

Marinari, Parisi, and Zuliani have studied the
Edwards-Anderson spin glass model at a field
$B=0.4$. They had previously studied the four-dimensional
case~\cite{MarinariParisi98d}
and in their comment~\cite{MarinariParisi99}
to our paper~\cite{HoudayerMartin99b}
they present results for
the $d=3$ case using the same techniques. The cornerstone
of their approach is an out-of-equilibrium
estimate of $q_{\mbox{min}}$. They find this quantity to be
different from the (equilibrium) mean value of $q$,
giving evidence for
replica symmetry breaking (RSB).
However we see a danger in relying on out-of-equilibrium measurements:
metastable states that do {\it not} contribute to the
(equilibrium) $P(q)$ (because they have excess free energies
diverging with system size) may very well
contribute to out-of-equilibrium overlaps.

Since we have some doubts about the validity of out-of-equilibrium
measurements of $q_{\mbox{min}}$, let us consider the evidence for
RSB in the presence of a field using {\it equilibrium} measurements.
Most work has been performed
in $d=4$ where there is a clear signal of a growing 
spin glass susceptibility. Less clear is whether this quantity
actually diverges at $T>0$ and $B>0$: fits are compatible with
such a divergence but it is difficult
to conclude that there is a {\it finite} temperature transition.
Measurements of
higher order cumulants of $P(q)$ are very disappointing: the
cumulants do not cross as they do in zero field, there
is no clear pattern in the data, and in fact
there is no sensible way to
extrapolate the data to larger sizes. 
Furthermore, $P(q)$ has a long tail at $q<0$ that cannot
be there in the thermodynamic limit, and 
there is no hint yet of a 
delta function peak at $q=q_{\mbox{min}}$. The situation in
$d=3$ is even more ambiguous because the evidence for a diverging
spin glass susceptibility is much weaker. 
Nevertheless, since simulations of
the Sherrington-Kirkpatrick model run into similar
difficulties, one need not conclude that such 
results disfavor a RSB scenario in finite dimensions.
But it is also fair to say that there is today no substantial evidence
via equilibrium measurements for an Almeida-Thouless
transition line in $d=3$.

\begin{figure}
\begin{center}
\resizebox{0.9\linewidth}{!}{\includegraphics{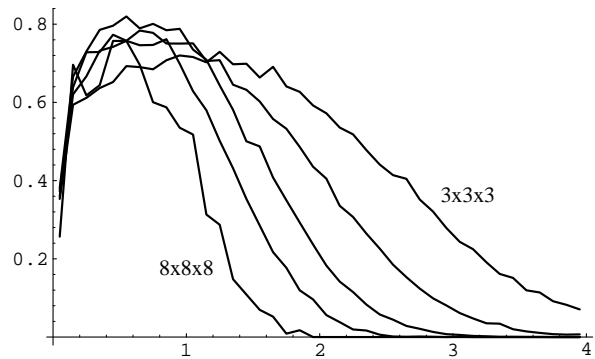}}
\end{center}
\caption{$r(N,B,0.15)/\sqrt{N}$ versus $B$.}
\label{fig1}
\end{figure}

Part of the difficulty stems from
the nearby critical point ($T_c, B=0$) that
leads to severe finite size effects; this may explain
the non crossing of the Binder cumulants for different
size lattices. Staying away from that critical point requires
lowering the temperature, leading to insurmountable
difficulties for thermalizing the lattices. Our approach
bypasses this problem by taking the zero temperature
limit and finding ground states. Doing so, we
found that the finite size
effects for the mean field model were very small
as shown in figure 3 of our paper. A comparison to what occurs
in simulations at finite temperature
of the Sherrington-Kirkpatrick model suggests that the ($T_c$, $B=0$)
critical point is quite far away from where we work. Extrapolating 
this to
the Edwards-Anderson model case, we are led to
conclude that the crossing points of the curves
of figure 1 in our paper simply converge to $B=B_c=0$.
We tried to substantiate this hypothesis by finite size scaling
(figure 2 in our paper). If on the other hand one insists on
having a critical value $B_c>0$ of the field, we are led to ask
whether 
the curves $r(N,B,0.15)/\sqrt{N}$ vs $B$ superpose
at large $N$ on a curve extending to $B>0$. Our
data is displayed in figure~\ref{fig1}. 
Judging from this figure, we probably would need to work with
lattices larger than $20^3$ before such behavior could
be seen. This may be feasible in
the not so distant future, but much remains to be done.
\bigskip

\noindent 
J. Houdayer and O.C. Martin\\
LPTMS, Universit\'e Paris-Sud, F-91405 Orsay, France

\medskip

\noindent
July 22, 1999


\bibliographystyle{prsty}
\bibliography{/remote/home/plato/houdayer/Papers/Biblio/references}

\end{document}